# Parallel optically detected magnetic resonance spectrometer for dozens of single nitrogen-vacancy centers using laser-spot lattice

*Mingcheng Cai\*, Zhongzhi Guo\*, Fazhan Shi#, Chunxing Li, Mengqi Wang, Wei Ji, Pengfei Wang#, Jiangfeng Du#*

Abstract: We develop a parallel optically detected magnetic resonance (PODMR) spectrometer to address, manipulate and read out an array of single nitrogen-vacancy (NV) centers in diamond in parallel. In this spectrometer, we use an array of micro-lens to generate 20 × 20 laser-spot lattice (LSL) on the objective focal plane, and then align the LSL with an array of single NV centers. The quantum states of NV centers are manipulated by a uniform microwave field from a Ω-shape coplanar coil. As an experimental demonstration, we observe 80 NV centers in the field of view. Among them, magnetic resonance (MR) spectrums and Rabi oscillations of 18 NV centers along the external magnetic field are measured in parallel. These results can be directly used to realize parallel quantum sensing and multiple times speedup compared with the confocal technique. Regarding the nanoscale MR technique, PODMR will be crucial for high throughput single molecular MR spectrum and imaging.

KEYWORDS:
Nitrogen vacancy center, diamond, optically detected magnetic resonance, laser-spot lattice, quantum sensing

The nitrogen-vacancy (NV) center in diamond is a highly promising sensor due to its long spin coherence time of millisecond[1] and spin-dependent fluorescence[2, 3]. It has been explored as an outstanding magnetic quantum sensor[4, 5] with single-spin sensitivity[6] and nanometer spatial resolution[7]. These advantages promise broad applications in biology and condensed matter physics[8-13]. Among these, single-molecule scale magnetic resonance spectroscopy is one of the most amazing research fields, which push the detectable sample volumes from ~ 100 μL to $10^{-21}$ L and could be applied to structural analysis and dynamic characterization of single molecules in chemistry and biology. Nanoscale nuclear magnetic resonance (NMR) spectrum[14-20], single-biomolecule electron spin resonance (ESR) spectrum[21, 22] and NMR spectrum[23] have been demonstrated in the NV center system.

The magnetic magnitude of a proton in the samples is only a few nano-tesla at 10 nm distance and will decrease by cube distance. Correspondingly, the signal of an electron is roughly a thousand times larger than a proton but still weak. To sense such a weak magnetic signal, a single NV center is located in close proximity to the samples and then collects the signal from the spins around the NV center in the range of tens of nanometer by dynamic decoupling sequences. A confocal microscope is used to address the NV centers one by one to detect a single molecule or nanometer-scale samples. It needs a long time to accumulate photons for enough signal-to-noise ratio (SNR), i.e., typically several hours to days, due to the weak magnetic signal of samples, low fluorescence emission and collection efficiency, and poor quantum coherent features of shallow NV centers. Considering the about 30 nm detection range of a single NV center and the random distribution of molecules, we usually have to check a few tens of single NV centers for valid signals[21, 22]. This leads to low efficiency in experiments and thus limits the further applications. An intuitive way to break the limitation is using as many single NV centers as possible in parallel to speed up the measurements. Although two previous works have shown the possibility using ensemble NV centers in pillar array[24, 25], ensemble NV centers are powerless on magnetic resonance spectroscopy at single-molecule scale. Besides, a strong power laser is needed due to the uniform optical illumination on the whole detection area.

In this paper, we develop a parallel optically detected magnetic resonance (PODMR) spectrometer for an array of single NV centers. To improve the laser efficiency, we use a micro-lens array to generate the laser-spot lattice (LSL) to pump dozens of single NV centers in parallel. All of the NV sensors are controlled by resonant microwave fields radiated from a uniform large-area Ω-shape coplanar coil. The fluorescence from single NV centers is collected respectively in different pixels of an Electron-Multiplying charge-coupled device (EMCCD). As a demonstration of our spectrometer, we generate 20×20 LSL and then measure the PODMR of 18 NV centers. In principle, hundreds and thousands of single NV centers can be used in parallel with larger focused laser spots lattice.

Figure 1 shows the optical part of the platform. The heart of the optical path is the micro-lens array laser modulation system to generate LSL. Comparing to the uniform wide-field exciting laser, the LSL achieves higher exciting efficiency and lower background. The LSL generated by the micro-lens array is stable enough for long time experiments and easy to be used without a complex optical system.

We use a micro-lens array, a convex lens $f_1$ and an objective lens sequentially to make the LSL pass through the diamond and focus on the diamond surface. The micro-lens array and the convex lens $f_1$ is placed with the collective focal plane and in front of the dichroic mirror. The interval of LSL can be adjusted by replacing the convex lens $f_1$ to match the NV array. The interval of the LSL on the diamond surface can be calculated by

$$d_s = \frac{d_a}{m} = \frac{d_a}{\frac{f_1}{f_t} \times M} \qquad (1)$$

where $d_a$ is the interval of the micro-lens array, $m$ is the magnification of the imaging system, $f_1$ is the focal length of the focus lens $f_1$, and the $f_t$ is the length of the tube length of the objective, $M$ is the magnification of the objective. For the interval $d_s = 2$ μm, we use $d_a$=150 μm, $f_1$=225 mm, $f_t$=180 mm, $M$=60. Our micro-lens array is a commercial lens (Thorlabs MLA150-7AR-M) with the filling factor, the interval and the focal length of 74.5%, 150 μm and 5.2 mm, respectively. The convex lens $f_1$ (Union optic JGS1) has a focal length of 225 mm. The distance between the lens $f_1$ and the objective is 150 mm and most of the excitation laser can enter the aperture of the objective.

We use a 532 nm continuous-wave green laser (CNI OEM-V-532-5W) with water-cooling (CWUL-05) to generate high power laser. Three AOMs (Isomet M1133-aQ80L-1.5) are used to switch the continuous laser to pulsed laser of microsecond and its on-off ratio is at least $10^7$:1. Between the laser and AOMs, a half-wave plate (Thorlabs WPH10M-532) rotates the laser to vertically polarization for the highest efficiency through AOMs. An expander (Thorlabs BE02-532) is used to adjust the diameter of the laser spot to cover about 20×20 micro-lens. A dichroic mirror (Thorlabs DMLP550L) reflects the laser to the objective lens (OLYMPUS LUCPlanFL N 60×) while transmitting the fluorescence from NV centers. The laser is focused on the diamond surface and the fluorescence of NV centers is collected by the same objective lens. An EMCCD (Princeton Instruments ProEM-HS:1024B×3) with a focused lens ($f_2$ = 200 mm, Thorlabs AC508-200-B-ML) is used to map the fluorescence from NV centers.

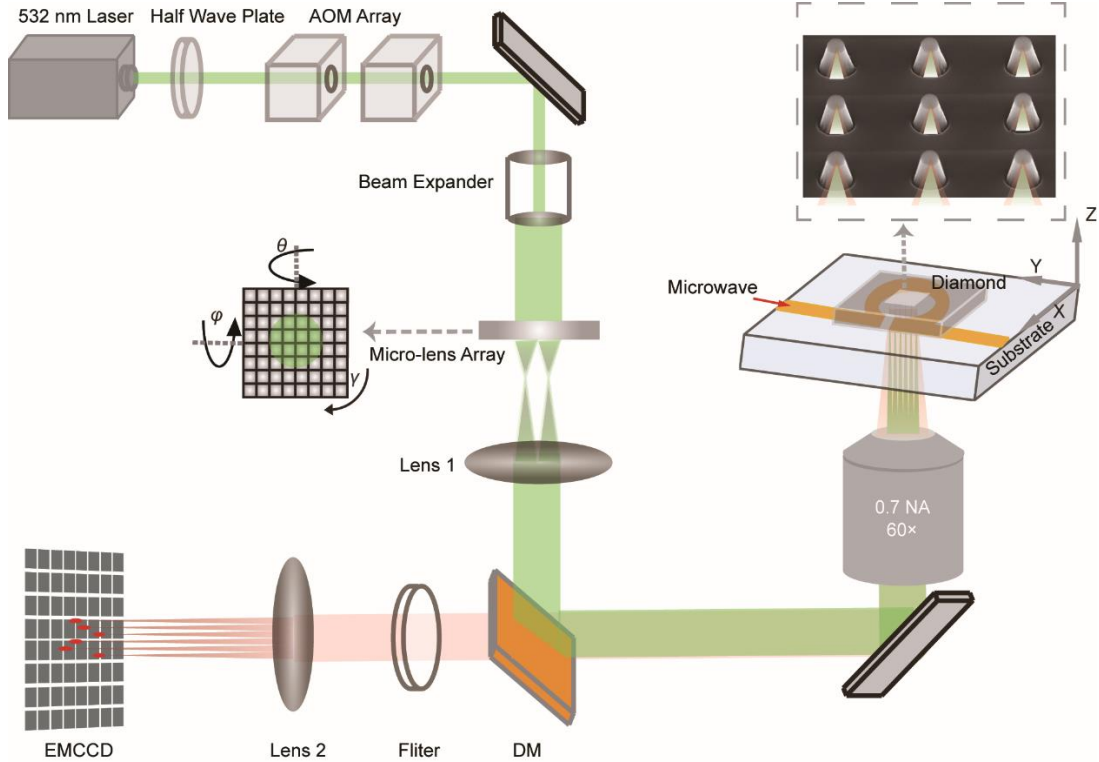

FIG. 1. Schematic view of the optical path. It consists of 532 nm laser excitation, a micro-lens array, an air objective, an EMCCD camera, and several of optical units.

We design 3 stages for rotating the micro-lens array and another 2 stages for translationally moving the diamond to align the LSL and the NV centers (Figure 1). We combine one electric goniometric stage (Newport M-GON65-U for angle $\varphi$) and two electric rotation stages (stage 1: Newport M-481-A for angle $\theta$ and stage 2: Thorlabs DDR25/M for angle $\gamma$) to rotate the micro-lens array. All the rotation axes are designed to coincide or intersect with the central axis of the optical path to ensure the pillar array and LSL always viewed in EMCCD during rotation. We use a micro positioner combined with a 3-dimension (3D) piezo scanner (Physik Instrumente P-562.3CD) of 200 μm range to precisely translate and align the pillar array to the LSL.

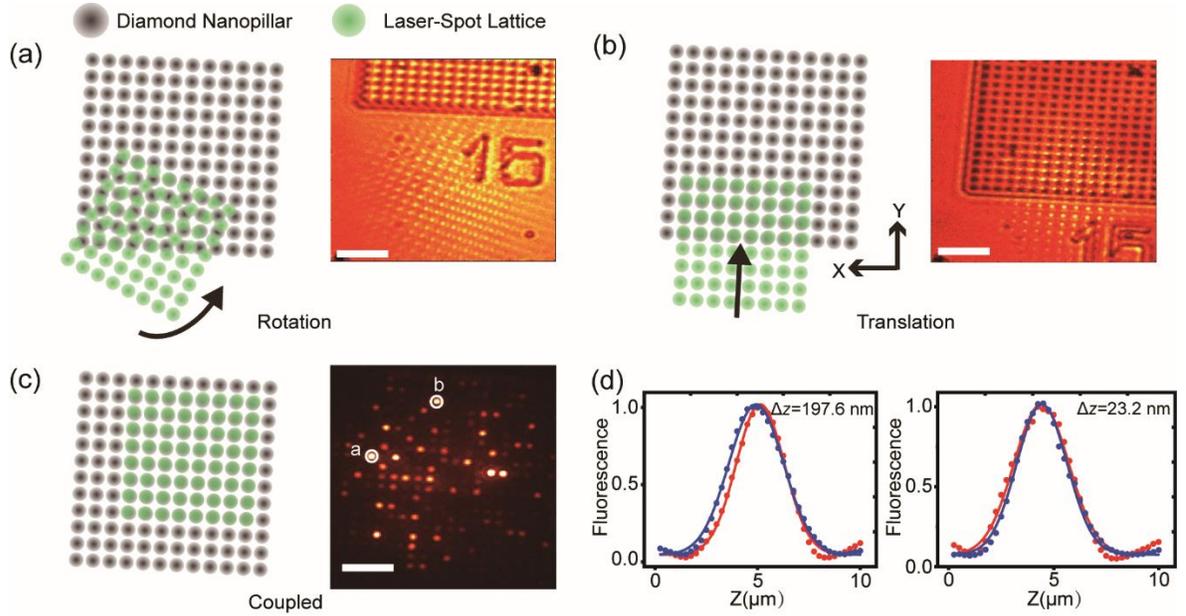

**Figure 2.** The alignment process of the LSL with diamond pillar array. (a) The sketch map (left) and picture (right) under a bright field microscopy of LSL and nanopillar array. At the very beginning, the laser array is neither focused on the diamond surface nor parallel with the nanopillar array. (b)The laser array is rotated to be parallel with the pillar array. (c) The laser array is well coupled with the NV array. The right figure shows the fluorescence map of the NV array after alignment. (d) The left normalized point spread function in the Z-direction of NV a (red dots) and NV b (blue dots) shows a large mismatch with $\Delta z$=197.6 nm. After delicate alignment of the space angle $\theta$ and $\varphi$ of the micro-lens array, the focused laser plane coincides with the NV array, i.e., $\Delta z$=23.2 nm. Scale bar in (a-c): 10 μm.

The process of aligning the laser spots to the NV array is as follows (Figure 2): first, move the pillar array into the view of EMCCD. Second, adjust the space angles $\theta$ and $\varphi$ of the micro-lens array. Third, rotate the laser array to the angle $\gamma$ until the pillar array and the laser array are parallel (from Figure 2(a) to Figure 2(b)). Finally, translate the pillar array by the piezo scanner precisely to make these two arrays align (from Figure 2(b) to Figure 2(c)). Whether the two arrays coincide can be decided by measuring the point spread functions of NV centers in different areas. We neglect the single NV centers with less fluorescence because of their location probably on the edge of the pillar. After that, when the two arrays coincide, all the remaining single NV centers locate in the same position. Figure 2(d) shows the deviation in the Z-axis of two NV centers when we modulate the space angle $\theta$ and $\varphi$ of the micro-lens array. We could confirm these two arrays coincide in Z-axis due to the deviation of 23.2 nm within the fitting error ~30 nm.

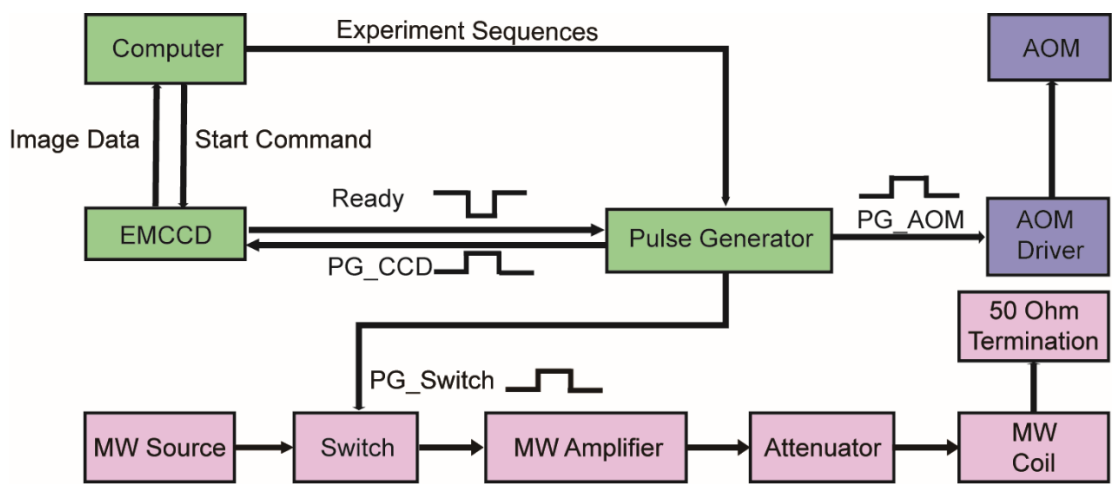

**Figure 3.** The microwave and electrical circuits. The workflow of one loop is as follows: first, the computer gives EMCCD a start command and EMCCD prepares to enter the state of waiting for an external trigger. Once EMCCD is ready, it gives a falling-edge trigger to the pulse generator (PG) and then PG starts running the pulse sequence to trigger AOMs, microwave, and EMCCD to do the experiment. Finally, after one experiment loop is done, the computer obtains an image from EMCCD and analyze it immediately. Then EMCCD automatically runs the cleaning cycle and then start the next loop.

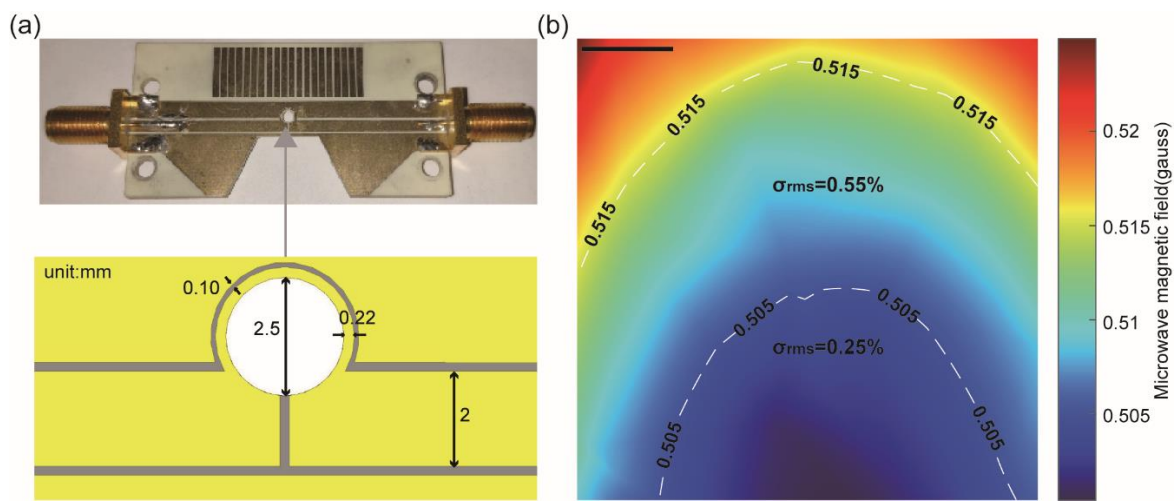

**Figure 4.** The Ω-shape microwave coplanar coil used for homogenous microwave delivery. (a) Fabricated Ω-shape coil and its detailed parameters. (b) Simulation of microwave magnetic field distribution of the coil. The diamond is roughly put in the center of the coil. The contour lines of magnetic field are drawn in the picture with their corresponding values. Every region is annotated with its root-mean-square inhomogeneity $\sigma_{rms}$. The bar in this figure is 100 μm.

Figure 3 shows the schematic view of the microwave system and the control process. A pin switch (Mini-Circuits ZASWA-2-50DRA+) is used to control the on and off of the microwave from the MW source (SG386 Stanford). The switch is controlled by transistor-transistor logic (TTL) signals from the Pulse Generator (Spincore PulseBlaster ESR-PRO-500, 500 MHz). Then the microwave is amplified by a high-power amplifier (Fairview Microwave SPA-060-50-SMA). An isolator blocks the microwave reflected by the MW coil.

We use an Ω-shape microwave coplanar coil to manipulate all the single NV centers in parallel. Figure 4(a) shows the picture and the structural details of the fabricated coil on the printed circuit board (PCB). The transmission characteristic of the Ω-shape coil is optimized with HFSS software simulation and the designed frequency range is 2-5 GHz. Figure 4(b) is the simulated distribution of the magnetic field on the surface of a 0.5 mm thick diamond mounted on the Ω-shape coil. And the input MW in the simulation is at the frequency of 2870 MHz and with a power of 1 W.

As shown in Figure 5, the 2D fluorescence of multiple single NV centers is projected onto an interest region of 260×251 pixels in our EMCCD camera. The NV centers are created by 10 keV $^{14}N^{2+}$ implantation in [100] face of an ultrapure bulk diamond ([N] < 5 ppb, ElementSix). The typical depth of the $^{14}N^{2+}$ ions was estimated to be 5 to 11 nm below the diamond surface. We fabricated trapezoidal cylinder-shaped nanopillars on the diamond. The detail of the fabrication of the diamond is described in Ref[26]. Our optical system yields an effective pixel size of ~200 nm with the field of view of 50×50 μm$^2$, which is the same size as our pillar array of 25×25 with an interval of 2 μm. There are totally 80 NV centers in the field of view. At each fluorescence spot, we take the pixel with maximum counts as the position of the single NV center. Then the computer program sums the counts of the 9 (3 × 3) pixels around the center as the fluorescence count of one single NV center corresponding to its Full Width Half Maximum (FWHM) of ~ 600 nm. The maximum fluorescence photon count of a single NV center is 130 k/s and the background is 10 k/s. Some NV centers appear very low fluorescence counts because they locate on the edge of the pillars thus resulting in a low photon collection efficiency.

To correct the shift caused by temperature instability and mechanical vibration, we use a 3D auto-alignment protocol. The protocol is based on maximizing the photon counts of NV centers by moving the diamond to realign the LSL and the NV array. Once the Nano-positioner moves a step, EMCCD takes an image of the fluorescence from multiple NV centers in this position. The computer program calculates the fluorescence intensity of all the selected NV centers in each image. Then all the fluorescence is summed and Gaussian fitted to find the position with maximum fluorescence. Finally, the diamond is moved relatively to maximize the fluorescence and continues the experiment.

For the following experiments of MR spectrum and Rabi Oscillations, the initialization and readout laser pulse is set to 1 μs which is optimized for the best contrast before experiments. We use an EM gain of 25 and the single sequence is repeated $5\times10^5$ times in each frame so that the maximum count is around half of the full well (~70 ke$^-$). The whole experiment progress is repeated 20 times to achieve a high signal to noise ratio (SNR).

The pulse sequence of MR spectrum measurement is shown in Figure 5(b). The MW frequency F varies from 2870 MHz to 2909 MHz in steps of 1 MHz, totally 40 data points. When the MW frequency is on resonance with the NV center, the NV center is rotated to $m_s=1$ under a MW pulse with the duration of 128 ns. Then the detected fluorescence decreases. Immediately after the signal frame, we use a frame with the same parameter but with no MW pulse (the dashed line rectangle) as the reference. We totally acquire 80 frames and the time consumption is 220 s.

We obtain MR spectrums of 18 NV centers with their quantum axis along the magnetic field and list 4 of them in Figure 5(d)-(g). The data is normalized by dividing the signal by the reference. As we set a static magnetic field of ~ 8 G along [111] diamond axis, the resonance frequencies are around 2892

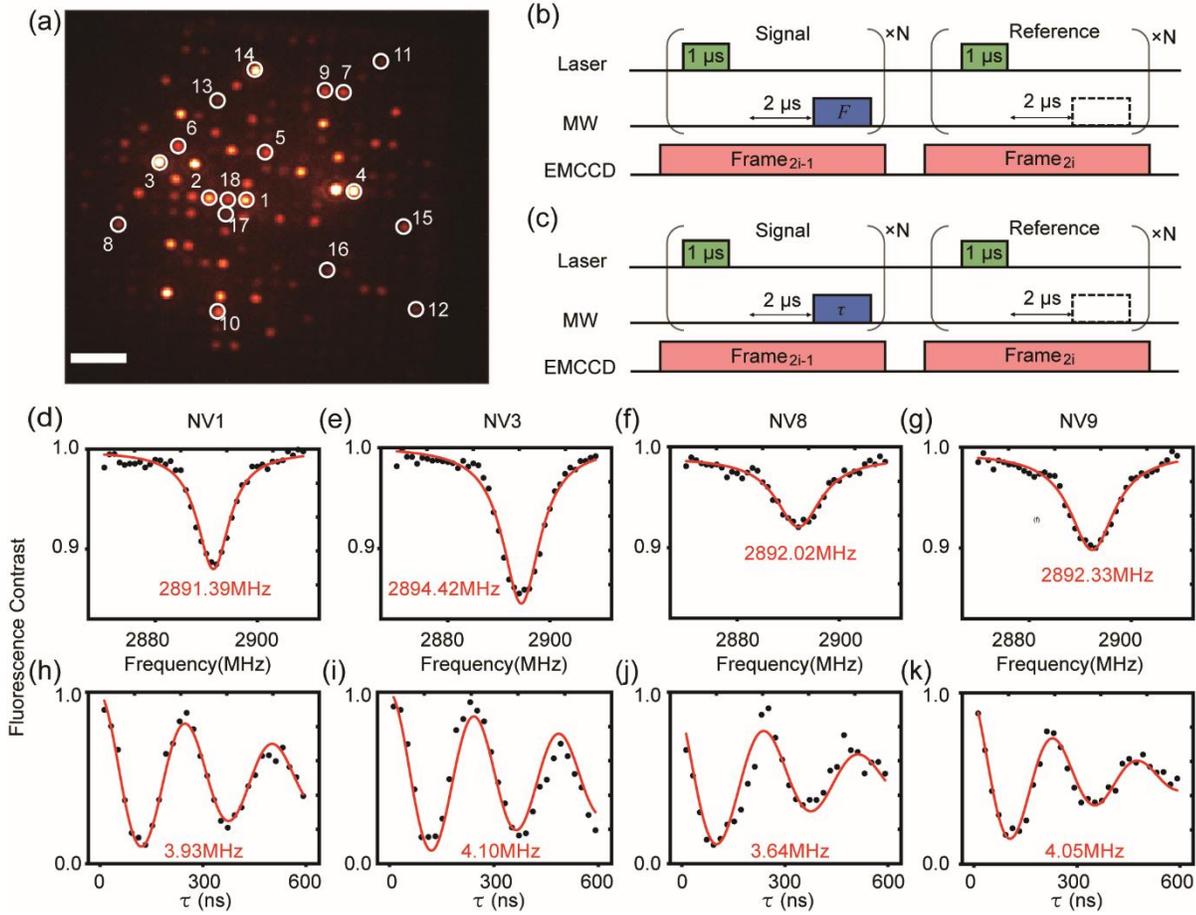

**Figure 5.** The parallel optically detected magnetic resonance experiment result. (a) The fluorescence map of NV centers. The bar in the figure is 5 μm. The NV centers along the magnetic field are marked with solid circles. (b) The pulse sequence for measuring MR spectrum. (c) The pulse sequence for measuring Rabi oscillations. (d)-(g) The normalized MR spectrum of four single NV centers (black dots) and the data is fitted by Lorentz function (red lines). (h)-(k) The corresponding Rabi oscillations of four single NV centers with $F$= 2892 MHz. The data is fitted by Sine-damp function (red lines). The errors in (d)-(k) are around 0.003.

MHz. The PODMR reveals the specific characters of NV centers locating in different nanopillars. The differences in resonance frequencies and contrasts of NV centers are due to the different local environments, such as magnetic inhomogeneity and local charge fluctuation[27].

The pulse sequence of Rabi oscillations measurement is shown in Figure 5(c). The MW frequency is fixed to $F$ = 2892 MHz. As the MW pulse duration $\tau$ varies, the NV spin rotates between $|0\rangle$ and $|1\rangle$. Immediately after the signal frame, we use a frame with the same parameter but with no MW pulse (the dashed line rectangle) as the reference. $\tau$ varies 30 times, from 12 ns to 592 ns in steps of 20 ns. We acquire 60 frames in 160 s.

As we use a single frequency MW pulse, 18 NV centers along the magnetic field are driven and appear clear Rabi oscillations. The normalized data results of the selected NV centers are shown in Figure 5(h)-(k). The Rabi frequency is calculated to be ~ 4 MHz. Considering the spatial uniformity of the microwave magnetic field, the variation of Rabi frequencies is mainly because of the off-resonance effect. Furthermore, it is able to drive all the 80 NV centers in all the 4 axes without crosstalk by using multitone MW.

In conclusion, we develop a PODMR platform with LSL. The platform has a MR measurement capacity for up to 20×20 NV centers and a homogeneous large-area MW magnetic field for uniform manipulation of all spins in the field of view. We observed 80 NV centers in the field of view, demonstrate MR spectrums and Rabi oscillations of 18 NV centers in diamond nanopillar array and observe the specific characters of each NV centers. This shows an 18 times efficiency improvement compared to laser scanning confocal microscopy. The demonstration shows a preliminary speed up comparing to the conventional ODMR and could be further improved for high throughput measurement. The efficiency of this work is mainly limited by the yield of single NV centers in pillar array (80 NV centers comparing with 400 spots in LSL).

The parallel platform vastly increases the efficiency of the experiments and ensures parallel detection of different points. With an achievable update of a larger micro-lens array and NV nanopillars, it is able to measure thousands of single NV centers in parallel. These improvements could make single-molecule scale magnetic resonance spectroscopy with single NV centers more applicable in medicine, biology and chemistry. For example, with the acceleration, the efficient single molecular MR spectroscopy of biological sample with short lives is visible by NV sensors. And our spectrometer could be combined with microfluidic and make the single NV center array more applicable for high-throughput[28] cellular measurement with single-molecule resolution.

## ASSOCIATED CONTENT

**Supporting Information**
The MR spectrums and Rabi oscillations of the remaining 14 NV centers (Figure S1). The details about our light path and the LSL (Figure S2-S4).

## AUTHOR INFORMATION

**Corresponding Author**
  **Fazhan Shi** – *Hefei National Laboratory for Physical Sciences at the Microscale and Department of Modern Physics, University of Science and Technology of China, Hefei 230026, China; CAS Key Laboratory of Microscale Magnetic Resonance, University of Science and Technology of China, Hefei 230026, China; Synergetic Innovation Center of Quantum Information and Quantum Physics, University of Science and Technology of China, Hefei 230026, China.*
  Email: fzshi@ustc.edu.cn
  **Pengfei Wang** – *Hefei National Laboratory for Physical Sciences at the Microscale and Department of Modern Physics, University of Science and Technology of China, Hefei 230026, China; CAS Key Laboratory of Microscale Magnetic Resonance, University of Science and Technology of China, Hefei 230026, China; Synergetic Innovation Center of Quantum Information and Quantum Physics, University of Science and Technology of China, Hefei 230026, China.*
  Email: wpf@ustc.edu.cn
  **Jiangfeng Du** – *Hefei National Laboratory for Physical Sciences at the Microscale and Department of Modern Physics, University of Science and Technology of China, Hefei 230026, China; CAS Key Laboratory of Microscale Magnetic Resonance, University of Science and Technology of China, Hefei 230026, China; Synergetic Innovation Center of Quantum Information and Quantum Physics, University of Science and Technology of China, Hefei 230026, China.*
  Email: qcmr@ustc.edu.cn
Authors
  **Mingcheng Cai** – *Hefei National Laboratory for Physical Sciences at the Microscale and Department of Modern Physics, University of Science and Technology of China, Hefei 230026, China; CAS Key Laboratory of Microscale Magnetic Resonance, University of Science and Technology of China, Hefei 230026, China; Synergetic Innovation Center of Quantum*

*Information and Quantum Physics, University of Science and Technology of China, Hefei 230026, China.*

**Zhongzhi Guo** – *Hefei National Laboratory for Physical Sciences at the Microscale and Department of Modern Physics, University of Science and Technology of China, Hefei 230026, China; CAS Key Laboratory of Microscale Magnetic Resonance, University of Science and Technology of China, Hefei 230026, China; Synergetic Innovation Center of Quantum Information and Quantum Physics, University of Science and Technology of China, Hefei 230026, China.*

**Chunxing Li** – *Hefei National Laboratory for Physical Sciences at the Microscale and Department of Modern Physics, University of Science and Technology of China, Hefei 230026, China; CAS Key Laboratory of Microscale Magnetic Resonance, University of Science and Technology of China, Hefei 230026, China; Synergetic Innovation Center of Quantum Information and Quantum Physics, University of Science and Technology of China, Hefei 230026, China.*

**Mengqi Wang** – *Hefei National Laboratory for Physical Sciences at the Microscale and Department of Modern Physics, University of Science and Technology of China, Hefei 230026, China; CAS Key Laboratory of Microscale Magnetic Resonance, University of Science and Technology of China, Hefei 230026, China; Synergetic Innovation Center of Quantum Information and Quantum Physics, University of Science and Technology of China, Hefei 230026, China.*

**Wei Ji** – *Laboratory of Interdisciplinary Research, Institute of Biophysics, Chinese Academy of Sciences, Beijing 100101, China*

**Author Contribution**
P.W., F.S., W.J. and J.D. conceived the idea. P.W., F.S. and J.D. guided the whole project. P.W., F.S. and M.C. designed the spectrometer. P.W., F.S., M.C. and Z.G. built the platform, analyzed the data, drew the figures, and composed the manuscript. M.W. fabricated the diamond sample. All authors have approved the final version of the manuscript.

## ACKNOWLEDGEMENTS

This work was supported by the National Key Research and Development Program of China (Grants No. 2016YFA0502400, 2018YFF01012500, 2018YFA0306600), the National Natural Science Foundation of China (Grants No. 81788101, 91636217, 11722544, 11761131011, 11874338, and 31971156), the CAS (Grants No. GJJSTD20200001, QYZDY-SSW-SLH004 and YIPA2015370), the Anhui Initiative in Quantum Information Technologies (Grant No. AHY050000), the national youth talent support program.

# Support Information

# Parallel optically detected magnetic resonance spectrometer for dozens of single nitrogen-vacancy centers using laser-spot lattice

*Mingcheng Cai\*[†,‡,§], Zhongzhi Guo\*[†,‡,§], Fazhan Shi[#,†,‡,§], Chunxing Li[†,‡,§], Mengqi Wang[†,‡,§], Wei Ji[⊥, ∥, §, ¶], Pengfei Wang[#,†,‡,§], Jiangfeng Du[#,†,‡,§]*

[†]Hefei National Laboratory for Physical Sciences at the Microscale and Department of Modern Physics, University of Science and Technology of China, Hefei 230026, China

[‡]CAS Key Laboratory of Microscale Magnetic Resonance, University of Science and Technology of China, Hefei 230026, China

[§]Synergetic Innovation Center of Quantum Information and Quantum Physics, University of Science and Technology of China, Hefei 230026, China

[⊥]National Laboratory of Biomacromolecules, CAS Center for Excellence in Biomacromolecules, Institute of Biophysics, Chinese Academy of Sciences, Beijing, China

[∥]Physical Science Laboratory, Huairou National Comprehensive Science Center, Beijing, China

[§]Center for Biological Instrument Development, Core Facility for Protein Research, Institute of Biophysics, Chinese Academy of Sciences, Beijing, China

[¶]College of Life Science, University of Chinese Academy of Sciences, Beijing, China

\*Both authors contributed equally to this work.

[#]Corresponding Authors

fzshi@ustc.edu.cn (Fazhan Shi)

wpf@ustc.edu.cn (Pengfei Wang)

qcmr@ustc.edu.cn (Jiangfeng Du)

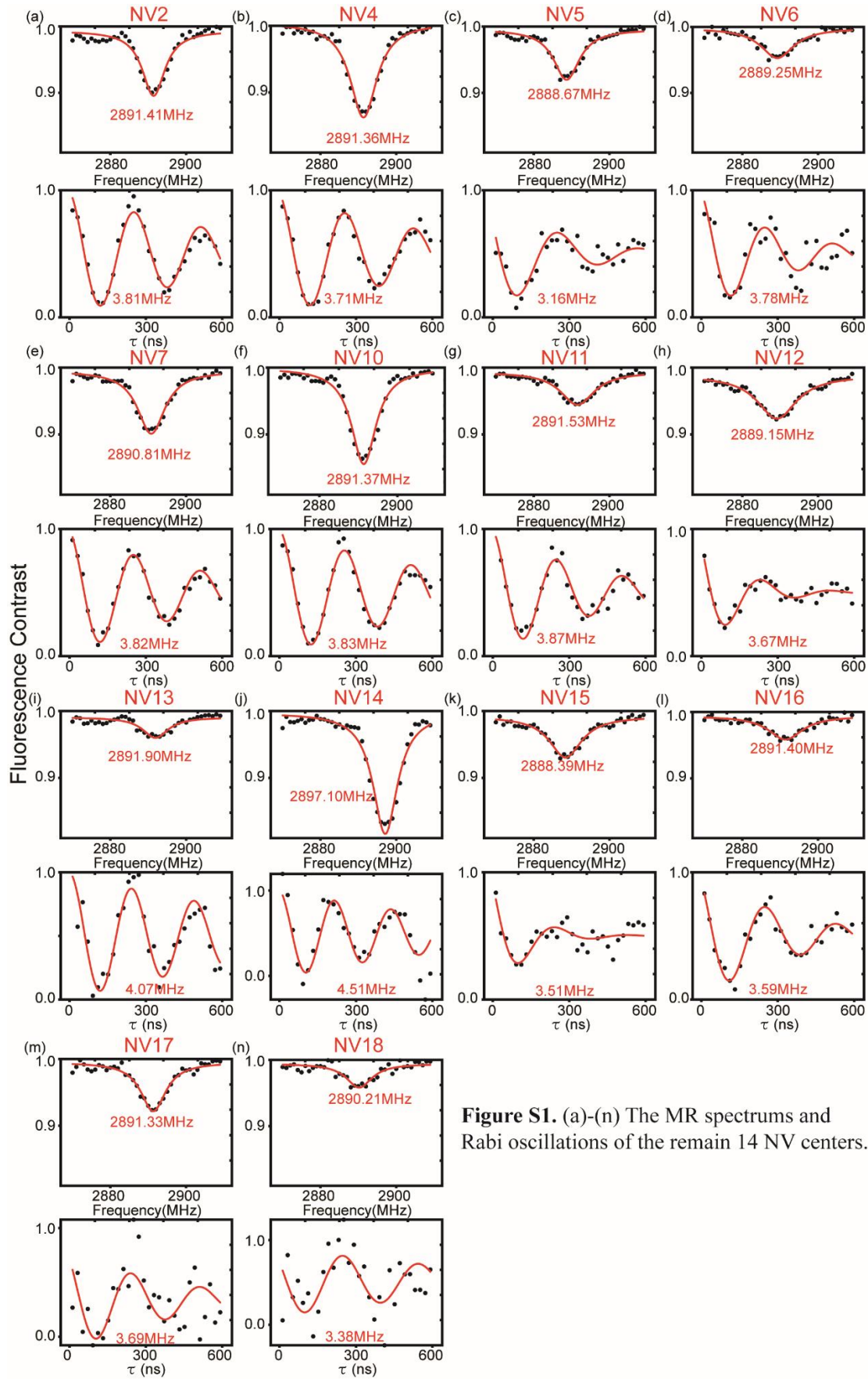

**Figure S1.** (a)-(n) The MR spectrums and Rabi oscillations of the remain 14 NV centers.

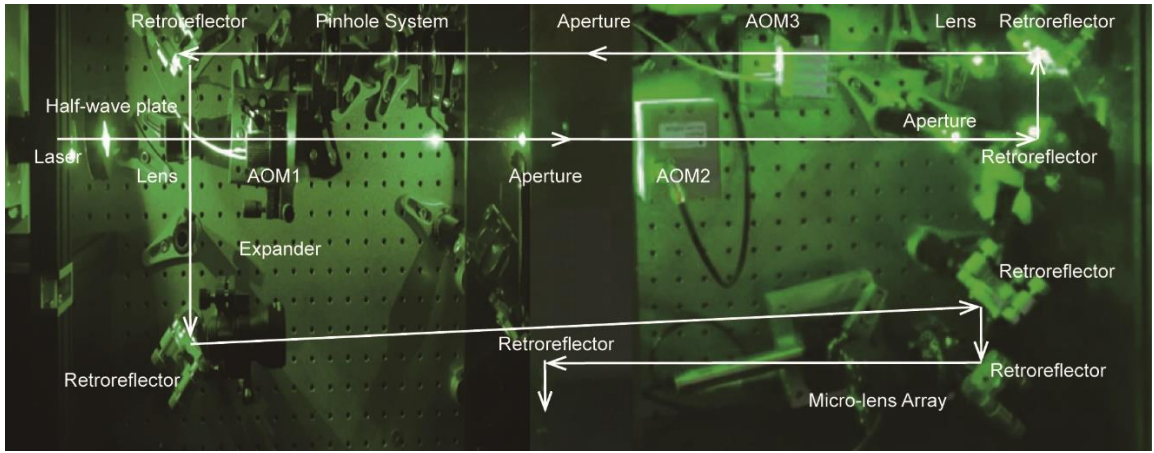

**Figure S2.** The light path of the exciting part of our spectrometer.

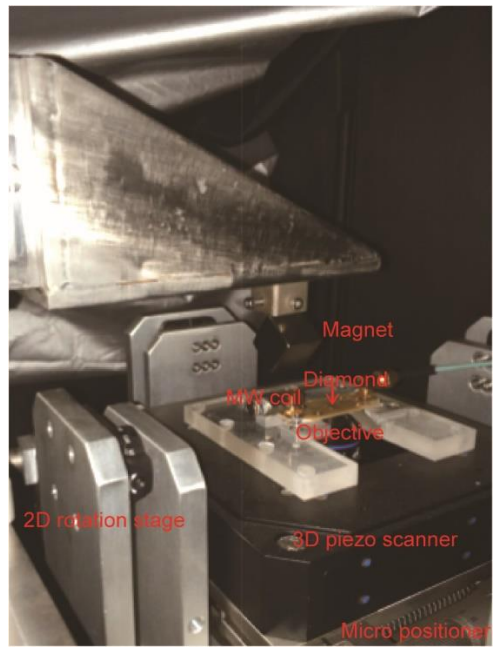

**Figure S3.** The sample part with its position and angle modulation system.

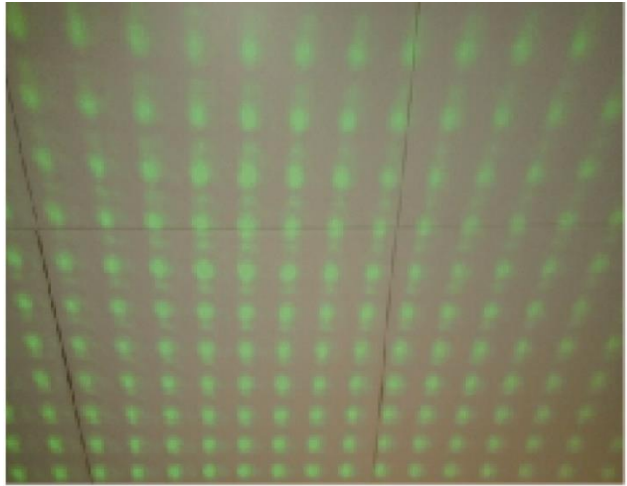

**Figure S4.** The laser-spot lattice (LSL) far (3 meter) behind the objective.